\input harvmac
\input psfig
\newcount\figno
\figno=0
\def\fig#1#2#3{
\par\begingroup\parindent=0pt\leftskip=1cm\rightskip=1cm\parindent=0pt
\global\advance\figno by 1
\midinsert
\epsfxsize=#3
\centerline{\epsfbox{#2}}
\vskip 12pt
{\bf Fig. \the\figno:} #1\par
\endinsert\endgroup\par
}
\def\figlabel#1{\xdef#1{\the\figno}}
\def\encadremath#1{\vbox{\hrule\hbox{\vrule\kern8pt\vbox{\kern8pt
\hbox{$\displaystyle #1$}\kern8pt}
\kern8pt\vrule}\hrule}}
\def\underarrow#1{\vbox{\ialign{##\crcr$\hfil\displaystyle
 {#1}\hfil$\crcr\noalign{\kern1pt\nointerlineskip}$\longrightarrow$\crcr}}}
%
\overfullrule=0pt

%

\def\bar{\overline}
\def\Z{{\bf Z}}

\def\S{{\bf S}}
\def\R{{\bf R}}

\font\zfont = cmss10 

\def\bigone{\hbox{1\kern -.23em {\rm l}}}
\def\ZZ{\hbox{\zfont Z\kern-.4emZ}}

\Title{hep-th/0007175}
{\vbox{\centerline{Overview Of $K$-Theory}
\bigskip
\centerline{ Applied To Strings}}}
\smallskip
\centerline{Edward Witten}
\smallskip
\centerline{\it School of Natural Sciences, Institute for Advanced Study}
\centerline{\it Olden Lane, Princeton, NJ 08540, USA}
\centerline{and}
\centerline{\it Department of Physics, Caltech, Pasadena CA 91125}
\centerline{and}
\centerline{\it CIT-USC Center For Theoretical Physics, Los Angeles CA}

\medskip

\noindent
$K$-theory provides a framework for classifying Ramond-Ramond (RR)
charges and fields.  $K$-theory of manifolds has a natural extension to
$K$-theory of noncommutative algebras, such as the algebras
considered in noncommutative Yang-Mills theory or in open string field
theory.  In a number of concrete problems, the $K$-theory analysis
proceeds most naturally if one starts out with an infinite set of
$D$-branes, reduced by tachyon condensation to a finite set.
This suggests that string field theory should be reconsidered for $N=\infty$.
\Date{August, 2000}
\newsec{Introduction And Definition Of $K(X)$}

A $D$-brane wrapped on a submanifold $S$ of spacetime may carry a nonzero
Ramond-Ramond (RR) charge.  RR fields are $p$-forms, and superficially it
seems that the conserved charge should be measured by the cohomology
class of the RR form (or of the cycle $S$ itself).
However, $D$-branes carry gauge fields; and gauge fields are not natural in
(co)homology theory.  They are natural in ``$K$-theory.''  
$K$-theory has been used to answer some questions about RR charges and fields;
the aim of the present article\foot{The article is based on my lecture at
Strings 2000, Univ. of Michigan, July 10, 2000 as well as earlier
lectures at the CIT-USC Center for Theoretical Physics.  
I thank both audiences for questions and comments.} 
is to give an overview
of this, along with some speculations (for which I have only very modest
evidence) about how one might want to rethink open
string field theory in the large
$N$ limit.  These subjects fill sections 1-3.  Some mathematical details
have been postponed to section 4.

\def\A{{\cal A}}

If $X$ is spacetime and $\A(X)$ is the commutative, associative
algebra of continuous complex-valued functions on $X$, then
the $K$-theory of $X$ can be defined in terms of representations of $\A(X)$.
A representation of a ring is usually called a module.
Here are some examples of $\A(X)$-modules.

The most obvious example of an $\A(X)$-module is $\A(X)$ itself.
For $f\in  \A(X)$ (regarded as a ring) and $g\in \A(X)$ (regarded
as a module), we define
\eqn\nobbo{f(g)=fg,}
where on the right hand side the multiplication occurs in $\A(X)$.
This obviously obeys the defining condition of a module, which is that
$(f_1f_2)(g)=f_1(f_2(g))$.

More generally, consider a $Dp$-brane (or a collection of $N$ $Dp$-branes
for some $N>0$) wrapped on a submanifold $S$ of $X$, with any Chan-Paton
gauge bundle $W$ on the $D$-brane.  Let $M(S)$ be the space of sections
of $W$, that is, the space of one-particle states for a charged scalar
coupled to the bundle $W$.  Then $M(S)$ is an $\A(X)$-module;
for $f\in \A(X)$, $g\in M(S)$, we simply set again $f(g)=fg$.
On the right hand side, the multiplication is defined by restricting $f$
(which is a function on $X$) to $S$ and then multiplying $f$ and $g$.

So in, say, Type IIB superstring theory, a collection of $D9$-branes
defines a representation or module $E$ of $\A(X)$.  A collection of
$\overline{D9}$-branes defines another module $F$.  So any configuration
of $D9$ and $\overline{D9}$-branes determines a pair $(E,F)$.

To classify $D$-brane charge, we want to classify pairs $(E,F)$ modulo
physical processes.  An important process \ref\sen{A. Sen, ``Tachyon
Condensation On The Brane-Antibrane System,'' JHEP {\bf 9808:012} (1998),
 hep-th/9805170.} is
brane-antibrane creation and annihilation -- the creation or annihilation
of a set of $D9$'s and $\overline{D9}$'s each bearing the same gauge bundle
$G$.  This amounts to
\eqn\eqrel{(E,F)\leftrightarrow(E\oplus G,F\oplus G).}
The equivalence classes make up a group called $K(X)$ (or $K(\A(X))$ if we
want to make the interpretation in terms of $\A(X)$-modules more explicit).
The addition law in this group is just
\eqn\nerel{(E,F)+(E'F')=(E\oplus E',F\oplus F').}
The inverse of $(E,F)$ is $(F,E)$; note that $(E,F)\oplus (F,E)
=(E\oplus F,E\oplus F)$, and using the equivalence relation \eqrel, this
is equivalent to zero.
$D$-branes of Type IIB carry conserved charges that take values in $K(X)$.  

\nref\witten{E. Witten, ``$D$-Branes And $K$-Theory,'' JHEP {\bf 9812:019} 
(1998), hep-th/9810188.}%
In the above definition of $K(X)$, we used only ninebranes,  even though,
as we explained earlier, an $\A(X)$ module can be constructed using
$Dp$-branes (or antibranes) for any $p$.  In fact, we can classify $D$-brane
charge just using the ninebranes, and then build the $Dp$-branes
of $p<9$ via pairs $(E,F)$ with a suitable tachyon condensate.
This construction (due to Atiyah, Bott, and Shapiro) was reviewed in
\witten, section 4, in the $D$-brane context.  
A more systematic explanation of the
definition of $K(X)$ in terms of modules, 
clarifying the role of ninebranes, can be found in 
section 4.

\bigskip\noindent
{\it Advantages Of $K$-Theory Description}

What do we gain by knowing that  $D$-brane charge is classified by
$K$-theory?

First of all, it is the right answer.  Wherever one looks closely at
topological properties of RR charges (or fields), one sees effects
that reflect the $K$-theory structure.  For example, there are stable $D$-brane
states (like the nonsupersymmetric $D0$-branes of Type I) that would
not exist if $D$-brane charge were classified by cohomology instead
of $K$-theory.  Conversely, it is possible to have a $D$-brane state that
would be stable if $D$-brane charge were measured by cohomology, but which
is in fact unstable (via a process that involves nucleation of $D9$-$\overline{
D9}$ pairs in an intermediate state).  This occurs \ref\dmw{E.
Diaconescu, G. Moore, and E. Witten, ``$E_8$ Gauge Theory, And
A Derivation Of $K$-Theory From $M$-Theory,'' hep-th/0005091.}
in Type II superstring
theory, in which a $D$-brane wrapped on a homologically nontrivial cycle
in spacetime is in fact in certain cases unstable.  Finally,
the $K$-theory interpretation of  $D$-branes is needed \ref\freed{D.
Freed and E. Witten, ``Anomalies In String Theory WIth $D$-Branes,''
hep-th/9907189.} to make sense of a certain global worldvolume anomaly.

But I think that there is a deeper reason that it is good to know about
the $K$-theory interpretation of $D$-branes: it may be naturally adapted
for stringy generalizations.  In fact (though some 
mathematical details have been postponed to
 section 4), we defined $K(X)$ in terms of
representations of the algebra $\A(X)$ of functions on spacetime.  
We can similarly define $K(\A)$ for any noncommutative algebra $\A$,
in terms of pairs $(E,F)$ of $\A$-modules.  By contrast, we would not
have an equally useful and convenient notion of ``cohomology'' if
the algebra of functions on spacetime is replaced by a  noncommutative ring.

For example, turning on a Neveu-Schwarz $B$-field, we can 
make $\A(X)$ noncommutative; the associated $K(\A)$ was used by Connes,
Douglas, and Schwarz in the original paper on noncommutative Yang-Mills theory
applied to string theory \ref\cds{A. Connes, M. Douglas, and A. Schwarz, 
``Noncommutative Geometry And Matrix Theory: Compactification On Tori,''
JHEP {\bf 9802:003} (1998), hep-th/9711162.}.  
This is an interesting example, even though
it involves only the zero modes of the strings.   One would much like
to have a fully  stringy version involving a noncommutative algebra constructed
using all of the modes of the string, not just the zero modes.

What is the right noncommutative algebra that uses all of the modes?
We do not know, of course.  One concrete candidate is the $*$-algebra of
open string field theory, defined in terms of gluing strings together. 
If I call this algebra $\A_{st}$, it seems plausible that $D$-brane
charge is naturally labeled by $K(\A_{st})$.  (For a manifold of
very large volume compared to the string scale, I would conjecture
that  $K(\A_{st})$ is the same as the ordinary $K(X)$ of topological
$K$-theory.)  I will come back to 
$K(\A_{st})$ in section 3.3.  First, I want to finish our survey of known 
applications of $K$-theory in string physics.

\newsec{$K$-Theory And RR Fields}

\nref\fh{D. S. Freed and M. J. Hopkins, ``On Ramond-Ramond Fields And
$K$-Theory,'' JHEP {\bf 0005:044} (2000), hep-th/0002027.}%
\nref\enw{E. Witten, ``Duality Relations Among Topological Effects
In String Theory,'' JHEP {\bf 0005:031} (2000), hep-th/9912086.}%
\nref\mw{G. Moore and E. Witten, ``Self-Duality, Ramond-Ramond Fields,
And $K$-Theory,'' JHEP {\bf 0005:032} (2000), hep-th/9912279.}%
 $K$-theory is relevant to understanding RR fields as well as charges
 \refs{\fh - \mw,\dmw}.

Naively speaking, an RR $p$-form field $G_p$ obeys a Dirac quantization
law according to which, for any $p$-cycle $U$ in spacetime,
\eqn\qnin{\int_U{G_p\over 2\pi}=\,{\rm integer}.}
If that were the right condition, then RR fields would be
classified by cohomology.

\nref\bersh{M. Bershadsky, C. Vafa, and V. Sadov, ``$D$-Branes And
Topological Field Theories,'' Nucl. Phys. {\bf B463} (1996) 420.}%
\nref\ghm{M. B. Green, J. A. Harvey, and G. Moore, ``$I$-Brane Inflow
And Anomalous Couplings On  $D$-Branes,'' Class. Quant. Grav. {\bf 14}
(1997) 47, HEP-TH/9605033.}%
\nref\chy{E. Cheung and Z. Yin, ``Anomalies, Branes, and Currents,'' Nucl.
Phys. {\bf B517} (1998) 69, hep-th/9710206.}%
\nref\mm{G. Moore and R. Minasian, ``$K$-Theory And Ramond-Ramond Charge,''
JHEP {\bf 9711:002} (1997), hep-th/9710230.}%
\nref\fluw{E. Witten, ``On Flux Quantization In $M$-Theory And The Effective
Action,'' J. Geom. Phys. {\bf 22} (1997) 1, hep-th/9609122.}%
But that is not the right answer, because the actual quantization
condition on RR periods is much more subtle than \qnin.  There are a variety
of corrections to \qnin\ that involve spacetime curvature and the gauge
fields on the brane, as well as self-duality and global anomalies
\refs{\bersh - \fluw,\enw,\mw,\dmw }.

The answer, for Type IIB superstrings,
 turns out to be that RR fields are classified by $K^1(X)$.
For our purposes, $K^1(X)$ can be defined as the group of components of
the group of continuous maps from $X$ to $U(N)$, for any sufficiently
large $N$.

This statement means that topological classes of RR fields on $X$ are classified
by a map $U:X\to U(N)$ for some large $N$.  The relation of
$G_p$ to $U$ is roughly $G_p\sim \Tr\,(U^{-1}dU)^p$; here I have ignored
corrections due to spacetime curvature and subtleties associated with
self-duality of RR fields.

\nref\horavag{M. Fabinger and P. Horava, ``Casimir Effect Between World-Branes
In Heterotic $M$-Theory,'' Nucl. Phys. {\bf B580} (2000) 243, hep-th/0002073.}%
The physical meaning of $U$ is not clear.  For Type IIA, the analog is
that RR fields are classified by a $U(N)$ gauge bundle (for some large
$N$) with connection $A$ and curvature $F_A$, the relation
being $G_p\sim \Tr\,F_A^{p/2}$.  The analog for $M$-theory involves
$E_8$ gauge bundles with connection \refs{\fluw,\horavag,\dmw}.  
Again, the physical meaning
of the $U(N)$ or $E_8$ gauge fields is not clear.

The value of using $K^1$ to classify RR fields of Type IIB is that
this gives a concise way to summarize the otherwise rather complicated
quantization conditions obeyed by the RR fields.  In addition, this
framework is useful in describing subtle phase factors that enter in the
RR partition function.  In hindsight, once it is
known that RR charges are classified by $K$-theory, one should have suspected
a similar classification for RR fields.  After all, RR charges produce
RR fields!  So the math used to classify RR charges must be similar to the math
used to classify RR fields.

Just like $K(X)$, $K^1(X)$ has an analog for any noncommutative
algebra $\A$.  Given $\A$, we let $\A_N$ denote the group of invertible
$N\times N$ matrices whose matrix elements are elements of $\A$.  Then
$K^1(\A)$ is the group of components of $\A_N$, for large $N$.

For example, for $\A=\A(X)$ the ring of complex-valued
continuous functions on $X$,
$\A_N$ is the group of maps of $X$ to $GL(N,{\bf C})$.  This is contractible
to the group of maps of $X$ to $U(N)$, so for large $N$
the group of components of $\A_N$
is the same as $K^1(X)$, as we defined it initially.

The existence of a generalization of $K^1(X)$ for noncommutative rings
means that the description of Type IIB RR fields by $K^1(X)$ in the long
distance limit may
be a useful starting point for stringy generalizations.

\newsec{$N\to\infty$}

In the last section, $N$ was a sufficiently large but finite integer.
Our next task will be to describe some things that  depend on setting
$N$ equal to infinity.

Before doing so, let us recall the role of the $N\to\infty$ limit
in physics.   It is important in the conjectured link of gauge theory
with strings; in the old matrix models that are used to give soluble
examples of string theory; in the matrix model of $M$-theory; and in the
correspondence between gravity in an asymptotically AdS spacetime
and conformal field theory on the boundary.

My theme here will be to suggest that we should somehow study
$D$-branes with $N=\infty$, with tachyon condensation to annihilate most
of the branes and reduce to something more manageable.
To motivate this, I will consider two concrete questions that seem
to require taking the number of $D$-branes to be infinite.
One question is the relation of Type IIA superstrings to $K$-theory;
the other is the inclusion of a topologically non-trivial NS three-form
field $H$ in the $K$-theory classification of RR charges and fields. 

After giving the talk, I became aware of a standard problem that
has already been interpreted in terms of an infinite number of $D9$ and
$\overline{D9}$-branes with tachyon condensation to something manageable
(see \ref\hori{
K. Hori, ``$D$-Branes, $T$-Duality, and Index Theory,''
hep-th/9902102.}, sections 2.2, 4, and 5).  This is a problem involving
a $D5$-brane probe of a $D5$-$D9$ system.

\subsec{Type IIA}

For Type IIB superstrings, we used $K(X)$ to classify RR charges, and
$K^1(X)$ to classify RR fields.

The $T$-dual statement is that for Type IIA, $K^1(X)$ should classify RR
charges, and $K(X)$ should classify RR fields.  (By Bott periodicity,
$K^{i+2}(X)=K^i(X)$, so the only $K$-groups of $X$ are $K^0(X)$, which we have
called simply $K(X)$, and $K^1(X)$.)

The most concrete and natural attempt to explain in general why
$K^1(X)$ classifies RR charges for Type IIA is that of Horava
\ref\horava{P. Horava, ``Type IIA $D$-Branes, $K$ Theory, and Matrix Theory,''
ATMP {\bf 2} (1999) 1373, hep-th/9812135.}.  
The starting point here is to consider
a system of $N$ unstable $D9$-branes of Type IIA.
The branes support a $U(N)$ gauge field and a tachyon field $T$ in the
adjoint representation of $U(N)$.  There is a symmetry $T\to -T$.

The effective  potential for the tachyon field is believed to have the general
form
\eqn\jinglo{V(T)={1\over g_{st}}\Tr\,F(T),}
where the function 
$F(T)$ is non-negative and, after scaling $T$ correctly, vanishes
precisely if $T=\pm 1$.  Hence $V(T)$ is minimized if and only if
every eigenvalue of
$T$ is $\pm 1$.  

It was argued in \horava\ that, in flat $\R^{10}$, one can  make
supersymmetric $Dp$-branes (for even $p$) as solitons of $T$.
For example, to make a $D6$-brane, we set $N=2$.  Let $\vec x$ be
the three coordinates normal to the $D6$-brane, and set
\eqn\turnon{T={\vec\sigma\cdot\vec x\over |x|}f(|x|),}
where $f(r)\sim r$ for small $r$, and $f(r)\to\infty$ for $r\to\infty$.
So for $|x|\to\infty$, the eigenvalues of $T$ are everywhere $\pm 1$.
Near $x=0$, there is a topological knot that we interpret as the $D6$-brane.

In flat ${\bf R}^{10}$, one can similarly make $Dp$-branes for other
even $p$.  But on a general spacetime, this does not work for arbitrary
$Dp$-branes unless we set $N=\infty$.
The problem  is most obvious if
$X$, or at least its spatial part, is compact. The tachyon field
$T$, being adjoint-valued, maps $X$ to the Lie algebra of $U(N)$; since
the Lie algebra is contractible, $T$
carries no topology.  So a map
from $X$ to the Lie algebra does not represent an element of $K^1(X)$;
indeed, it does not carry topological information at all. 

To define an element of $K^1(X)$, we need the group, not the Lie algebra;
a map $U:X\to U(N)$ does the job, as I have stated before.

\def\U{{\cal U}}
\def\H{{\cal H}}
Amazingly, as Atiyah and Singer showed long ago, we get back the right
topology from the Lie algebra if we set $N=\infty$!
We have to interpret $U(\infty)$ to be the unitary group $\U$ of a Hilbert
space $\H$ of countably infinite dimension. (Such a Hilbert space is
also called a separable Hilbert space.)  We interpret the $N=\infty$
analog of the space of hermitian $N\times N$ matrices
to be  the space of bounded self-adjoint operators $T$ on $\H$
whose spectrum is
as follows: there are infinitely many positive eigenvalues and infinitely
many negative ones, and zero is not an accumulation point of the spectrum.
(The last condition makes $T$ a ``Fredholm operator.'')
Physically, $T$ should be required to obey these conditions, since
they are needed to make the energy and the $D8$-brane charge finite.  In fact,
to make the energy finite, almost all the eigenvalues of $T$ are very close
to $\pm 1$.
Anyway, with these
conditions imposed on $T$, 
it turns out that the space of $T$'s has the same topology
as that of $U(N)$ for large $N$.

So we can use  tachyon condensation on a system of $D9$-branes to describe
RR charges for Type IIA.  But we have to start with infinitely many 
$D9$-branes, which then undergo tachyon condensation down to a configuration
of finite energy.  

\bigskip\noindent{\it Obstruction To Finite $N$}

Let us now explain in more concrete terms the obstruction to making
$D$-branes in this way for finite $N$, and how it vanishes for $N=\infty$.

Let us go back to the example of a $D6$-brane constructed with 2 $D9$-branes.
We took the transverse directions
to be a copy of $\R^3$, and the  tachyon field to be
\eqn\ucx{T={\vec\sigma\cdot\vec x\over |x|}f(|x|).}
If we try to compactify the transverse directions to $\S^3$, we run into
trouble because $T$ is not constant at infinity.  The conjugacy class
of $T$ is constant at infinity -- the eigenvalues of $T$ are everywhere
$1$ and $-1$ -- but $T$ itself is  not constant.

\def\L{{\cal L}}
Moreover, $T$ is not homotopic to a constant at infinity.  
If $T$ were homotopic to a constant near infinity, we would deform it to
be constant and then extend it over $\S^3$.  But it is not homotopic to a 
constant.

The basic 
obstruction to making $T$ constant at infinity is the ``magnetic charge.''
Let $\S^2$ be a sphere at infinity in $\R^3$. 
Over $\S^2$, we can define a line
bundle $\L_+$ whose fiber is the $+1$ eigenspace of $T$, and a line bundle
$\L_-$ whose fiber is the $-1$ eigenspace of $T$.  The line bundles $\L_+$
and $\L_-$ are topologically nontrivial -- their first Chern classes are
respectively 1 and $-1$.  As long as we try to deform $T$ preserving the
fact that its eigenvalues are  1 and $-1$, the line bundles $\L_\pm$ are
well-defined, and their first Chern classes are invariant.  So the nontriviality
of $\L_+$ (or $\L_-$) prevents us from making a homotopy to constant $T$.

Let us add some additional ``spectator'' $D9$-branes,  and see if anything
changes.  Suppose there are $M=2k$ additional branes, so that the total
number of branes is $N=2+M=2+2k$.  Let the tachyon
field be $T'=T\oplus U$, where $T$ is as above and $U$ is the sum of $k$
copies of the matrix
\eqn\pollygo{\left(\matrix{1 & 0 \cr 0 & -1\cr}\right)}
acting on the $2k$ additional branes.  

Thus $T'$ has near infinity
$k+1$ eigenvalues $+1$ and $k+1$ eigenvalues $-1$.  The $+1 $ eigenspace of
$T'$ is a vector bundle $V_+$ of first Chern class $1$ (since it is constructed
by adding a trivial bundle to $\L_+$), and the $-1$ eigenspace
of $T'$ is similarly a vector bundle $V_-$ of first Chern class $-1$.  
In particular,
$V_+$ and $V_-$ are nontrivial, so we have not gained anything by adding
the spectator branes:
$T'$ is not homotopic to a constant, and cannot be extended over infinity.
The nontriviality of $V_+$ is controlled by $\pi_1(U(k+1))=\Z$, which is
associated with the existence of a first Chern class.

Instead, what happens if we set $k=\infty$?  To be more precise, we take
the number of spectator
$D9$-branes to be countably infinite, and assume $T'=T\oplus U$, where $U$
is the direct sum of countably many copies of the matrix in \pollygo.  
We can still define the bundles
$V_+$ and $V_-$; their fibers are separable Hilbert spaces 
(that is, Hilbert spaces of countably infinite dimension).
  $U(k+1)$ is replaced by $\U$, the unitary group of a separable Hilbert space.
Now we run into the fundamental fact (Kuiper's theorem) that $\U$
is contractible; its homotopy groups are all zero.  Thus, any bundle
of separable Hilbert spaces is trivial.  In particular, $V_+$ and $V_-$
are trivial, so $T'$ is homotopic to a constant and can be extended over
infinity.  

So if the total number of unstable $D9$-branes is $N=\infty$,
we can make a $D6$-brane localized at a point in $\S^3$. More
generally, in view of 
the result of Atiyah and Singer, we can starting at $N=\infty$
build an arbitrary class in $K^1(X)$ via tachyon condensation.

In terms of applying this result to physics, there are a few issues
that we should worry about.  One question is simply whether it is physically
sensible to start with infinitely many branes and rely on tachyon condensation
to get us down to something of finite energy.  We will have to leave this
question to the future (but see \hori\ for an example).

Quite a different question is whether the answer that we have obtained
by setting $N=\infty$
is the right one for physics.   In the field of a $D6$-brane that is
localized at a point on $\S^3$, the equation for the RR  two-form field
$G_2$ (of which the $D6$-brane is a magnetic source) has no solution, since
``the flux has nowhere to go.''  

It seems that the situation is that $N=\infty$ corresponds to the
correct answer in classical open string theory, where the effective
action comes from worldsheets with the topology of a disc.  The RR fields
enter as a correction of relative
order $g_{st}$ (the closed string coupling constant)
coming from worldsheets with cylinder topology, and should be ignored in the
classical approximation.

The classification of $D$-branes by brane creation and annihilation
holds at $g_{st}=0$, and (as we recalled in section 1)
leads for Type IIB to a classification of
$D$-brane charge by $K(X)$.  To get the analogous answer -- namely
$K^1(X)$ -- for Type IIA   via unstable $D9$-branes and tachyon condensation, 
 we need to start at $N=\infty$.

Intuitively, in the absence of tachyon condensation, $N=\infty$ should
correspond to $g_{st}=0$, since the effective expanstion parameter for
open strings is $g_{st}N$. If $N$ is infinite, then prior to tachyon
condensation, $g_{st}$ must be zero,
or the quantum corrections diverge.  If we want $g_{st}$ to be nonzero,
we need tachyon condensation to reduce to an effective finite value of $N$. 

\subsec{Turning On An $H$-Field}

A somewhat analogous problem is to consider $D$-branes when the
Neveu-Schwarz three-form field $H$ is topologically nontrivial.
We will carry out this discussion in Type IIB (for Type IIA, we would
have to combine what follows with what we said above in the absence
of the $H$-field).

Just as at $H=0$, we would like to classify $D$-brane states
by pairs $(E,F)$ (where $E$ is a $D9$ state and $F$ is a $\overline{D9}$ 
state) subject to the usual sort of equivalence relation.  
But there is a problem in having a $D9$ state in the presence of an $H$-field.

In fact, when $H$ is topologically non-trivial, one cannot have a single
$D9$-brane.  On the $D9$-brane, there is a $U(1)$ gauge field with
field strength $F$.  The relation $dF=H$ shows, at the level of
de Rham cohomology, that $H$ must be topologically trivial if a single
$D9$-brane is present.  This conclusion actually holds precisely,
not just in de Rham cohomology.

There is a special case in which there is a comparatively elementary
cure for this difficulty (see \witten, section 5.3 and \ref\kap{A. Kapustin,
``$D$-Branes In A Topologically Nontrivial $B$-Field,'' hep-th/9909089.}).  
If $H$ is torsion, that is if there is
an integer $M>0$ such that $MH$ is topologically trivial, then it is 
possible to have a set of $M$ $D9$-branes whose ``gauge bundle'' actually
has structure group $U(M)/\Z_M$, rather than $U(M)$. (The obstruction to
lifting the $U(M)/\Z_M$ bundle to a $U(M)$ bundle is determined by $H$.)
We will call such
a gauge bundle a twisted bundle. More generally,
for any positive integer $m$, we can have $N=mM$ $D9$-branes with
the structure group of the  bundle being $U(mM)/\Z_M$.
In such a situation, $D$-brane charge is classified, as one would guess,
by pairs $(E,F)$ of twisted bundles (or $D9$ and $\overline{D9}$ states)
subject to the usual equivalence relation.  The equivalence classes
make a group $K_H(X)$.

If one wishes to interpret $K_H(X)$ as the $K$-theory of representations of
an algebra, one must pick a particular twisted bundle $W$ and
consider a $D$-brane state with boundary conditions determined by 
 $W$.  The $W$-$W$ open strings transform in the adjoint 
representation, so the gauge parameters of the zero mode sector of the
open strings are
sections of $W\otimes \bar W$.  Notice that although $W$ is a twisted
bundle (with structure group $U(M)/\Z_M$ rather than $U(M)$),  $W\otimes
\bar W$ is an ordinary bundle, since the center acts trivially in the adjoint
representation.  

The sections of $W\otimes \bar W$ form an algebra, defined as follows:
if $s^i{}_j$ and
$t^k{}_l$ are sections of $W\otimes \bar W$ -- where the upper and lower
indices are respectively $W$- and $\bar W$-valued -- then their product
is $(st)^i{}_l=\sum_ks^i{}_kt^k{}_l$.
This is the algebra
 $\A_W(X)$  of all endomorphisms or linear 
transformations of the bundle $W$.  
The algebra of open string field theory, for $W$-$W$ open strings,
reduces to $\A_W$ if one looks only at the zero modes of the strings.
This is a sensible approximation at low energies in a limit in which
$X$ is very large compared to the string scale.

If $H$ is zero and
$W$ is a trivial rank one complex bundle, then $\A_W(X)$ is our
friend $\A(X)$.  If $H$ is zero and
$W$ is a trivial rank $N$ complex bundle, then
including $W$ means simply that there are $N\times N$ Chan-Paton matrices
everywhere.
So in this case,
 $\A_W(X)=\A(X)\otimes M_N$, where $M_N$ is the algebra of
$N\times N$ complex-valued matrices.  In general, whatever $H$ is,
$W$ is always trivial
locally, so locally $\A_W(X)$ is isomorphic to  $\A(X)\otimes M_N$.

A twisted bundle is equivalent to an $\A_W$-module, and the group $K_H(X)$
of pairs $(E,F)$ of twisted bundles (modulo the usual    equivalence)
coincides with
 $K(\A_W)$,
the $K$-group of
$\A_W$-modules.  This will be explained in section 4.
This assertion leads to an immediate puzzle; $K_H(X)$ as defined in terms
of pairs $(E,F)$ of twisted bundles is manifestly independent of $W$
while $K(\A_W)$ appears to depend on $W$.  Indeed, given any two distinct
twisted bundles $W$ and $W'$, the corresponding algebras $\A_W$ and 
$\A_{W'}$ are distinct, but are ``Morita-equivalent.''  This concept
is explained in section 4, where we also show that the Morita equivalence
implies that $K(\A_W)=K(\A_{W'})$.

So far, we have only considered the case that $H$ is torsion.
A typical example, important in the AdS/CFT correspondence,
 is the spacetime $X={\rm AdS}_5\times {\rm RP}^5$,  where
a torsion $H$-field on ${\rm RP}^5$ is used to describe $Sp(n)$ rather
than $SO(2n)$ gauge theory in the boundary CFT.

However, in most physical applications, $H$ is not torsion.  In that case,
we must somehow take a large $M$ limit of what has been said above.
The right way to do this has been shown by Bouwknegt and Mathai
\ref\boma{P. Bouwknegt and V. Mathai, ``$D$-Branes, $B$ Fields, And Twisted
$K$ Theory,'' JHEP  {\bf 0003:007} (2000), hep-th/0002023.} 
and Atiyah and Segal 
\ref\atse{M. F. Atiyah and G. B. Segal, unpublished.}.
The suitable large $M$ limit of $U(M)/\Z_M$ 
is  $PU(\H)=U(\H)/U(1)$.  In other words, for $M=\infty$, one replaces
$U(M)$ by the unitary group $U(\H)$ of a separable Hilbert space $\H$;
and one replaces $\Z_M$ by $U(1)$.
This means, in particular, that when $H$ is not torsion, one cannot have
a finite set of $D9$- or $\overline{D9}$-branes, 
but one can have an infinite set, with a suitable infinite rank
twisted gauge bundle $E$ or $F$.  Then $D$-brane charge is classified
by the group $K_H$ of pairs $(E,F)$ modulo the usual equivalence relation. 
A detailed explanation can be found in \boma. Kuiper's theorem
-- the contractibility of $\U=U(\H)$ -- plays an important role, as it did
in section 3.1.

This construction, in the $M=\infty$ limit, has the beautiful property,
explained in \boma, that the noncommutative algebra whose $K$-group
is $K_H$ is {\it unique}, independent of any arbitrary choice of twisted
bundle $W$ or $W'$.  This really depends on the number of $D9$ and $\overline
{D9}$ branes being infinite.  

\subsec{Stringy Generalization?}

At the risk of going out on a somewhat shaky limb, I will now try to
propose a stringy generalization of some of this.

We start with a closed string background and try to form an open string
algebra.  To do so, we must pick an open string boundary condition
-- call it $\alpha$.  Then, in open string field theory,
the $\alpha$-$\alpha$ open strings form an algebra $\A_\alpha$.  

This algebra is not unique; we could instead pick another boundary condition
$\beta$ and define another algebra $\A_\beta$.

The example of strings in a background $H$-field that is torsion,
and the associated algebras $\A_W$, $\A_{W'}$, suggests the
nature of the relation between $\A_\alpha$ and $\A_\beta$: they
are different algebras, but are Morita-equivalent and hence have the
same $K$-theory.

The assertion that $\A_\alpha\not=\A_\beta$, if true, is quite troublesome.
It is a sharp statement of the lack of manifest background independence
of open string field theory.   (Background independence in this context
means independence of the open string background; of course, classical 
open string field theory depends on a closed string background in which the
open strings propagate.)  It means that the formalism depends on which
open string background one uses in setting up the theory.

The example of open strings in an $H$-field suggests a cure: take the number of
$D9$- and $\overline{D9}$-branes to be infinite.

\def\K{{\cal K}}
Let $m\alpha$ and $m\beta$ denote $m$ copies of $\alpha$ or $\beta$
(that is, $\alpha$ or $\beta$ supplemented with $U(m)$-valued Chan-Paton
factors).  Then the conjecture is that $\A_{m\alpha}\not=\A_{m\beta}$
for any finite $m$, but that they are equal for $m=\infty$.  
Actually, the statement about what happens for $m=\infty$
can be formulated  much more precisely by analogy
with statements in \boma.  Let $\K$ be the algebra of compact operators in
a separable Hilbert space $\H$.  Then the conjecture is that
for any $\alpha$ and $\beta$,
\eqn\mukky{\A_\alpha\otimes \K=\A_\beta\otimes \K.}

The main evidence for the conjecture is that, according to \boma, the
corresponding statement ($\A_W\not=\A_{W'}$ but $\A_W\otimes\K=\A_{W'}\otimes
\K$) holds for the zero mode algebra in the presence of a torsion $H$-field.
In addition, if it is necessary to tensor with $\K$ before the algebras
become isomorphic, this helps explain why background independence in
open string field theory is so hard to understand; in other words
it helps explain why the theory
constructed with classical solution $\alpha$ and algebra $\A_\alpha$
looks different from the theory constructed with classical solution $\beta$
and algebra $\A_\beta$.  However,  the evidence for the conjecture is
quite limited; it may be that whenever $\alpha$ is continuously connected
to $\beta$, $\A_\alpha=\A_\beta$, and that the difficulty in understanding
background independence in open string field theory is ``just''
 a technical problem.

If the conjecture is right, one would suspect that to get a greater degree
of background independence in open string field theory, we should start
with infinitely many $D9$-branes and rely on tachyon condensation to get
us down to something reasonable.

For this to be useful, we would need a description of $\A_\alpha\otimes\K$
much simpler and more incisive than any description we have today for
$\A_\alpha$ or $\A_\beta$.  In such a hypothetical new description, the BRST
operator $Q$ might be harder to describe.  If that happened, we would
have to take our lumps!

\newsec{Mathematical Details}

This concluding section will be devoted to explaining a few of the mathematical
points that we have skimmed over so far.

Let $\A$ be a ring -- in fact, let us momentarily assume that
$\A$ is commutative and associative.  An $\A$-module $M$ is called
``free'' if it is a direct sum of copies of $\A$: $M=\A\oplus \A\oplus
\dots\oplus \A$.  An $\A$-module $M$ is called ``projective'' if there
is another $\A$-module $M'$ such that $M\oplus M'$ is free.

Let us consider what these definitions mean in case $\A=\A(X)$ is the
ring of continuous complex-valued functions on a manifold $X$.  We can
think of $\A$ as the space of sections of a trivial complex line bundle
${\cal O}$.  Given a trivial rank $n$ complex vector bundle
$V={\cal O}\oplus {\cal O}\oplus \dots\oplus{\cal O}$, the space
of sections of $V$ is a free module $M$ which is the sum of $n$ copies
of ${\cal A}$.

Now suppose we are given {\it any} complex vector bundle $E$ over $X$.
The space of sections of $E$ is an $\A$-module $M(E)$ (given a function
$a\in \A(X)$ and a section $m$ of $E$, we simply define $a(m)$ to be the
product $am$).\foot{In the informal spirit of our discussion, 
I have generally used the same
name $E$ for a bundle and the corresponding module.  In this paragraph
only, I distinguish $E$ from $M(E)$ in the notation, to facilitate the
statement of the Serre-Swan theorem.}  
$M(E)$ is a free module if and only if $E$ is a trivial
vector bundle.  But $M(E)$ is always projective.  Indeed, there is always
a ``complementary'' vector bundle $F$ such that $E\oplus F$ is trivial,
and hence $M(E)\oplus M(F)=M(E\oplus F)$ is free.  Conversely (by
a theorem of Serre and Swan), every projective $\A(X)$-module 
is $M(E)$ for some complex vector bundle $E$ over $X$.

For any ring $\A$, the projective modules form a semigroup: if $E$ and
$E'$ are projective modules, so is $E\oplus E'$.  Given any semigroup
(with an addition operation that we will write as $\oplus$),
there is a canonical way to form a group.  This is done the
same way that one builds the group of integers, starting with
the semigroup of positive integers.  An element of the group is a pair
$(E,F)$, with $E$ and $F$ elements of the semigroup (so in our situation,
$E$ and $F$ are projective modules) and subject to the equivalence
relation $(E,F)\cong (E\oplus G,F\oplus G)$ for any element $G$ in the
semigroup.  Pairs are added by $(E,F)+(E',F')=(E\oplus E',F\oplus F')$.
The equivalence classes form a group (the zero element or
additive identity
is $0=(G,G)$ for any $G$, and the additive inverse of $(E,F)$ is $(F,E)$,
since $(E,F)+(F,E)=(E\oplus F,E\oplus F)=0$).
In case one starts with the semigroup of projective modules for a ring
$\A$, the group formed this way is called $K(\A)$.

For $\A=\A(X)$, we have in section 1
defined $K(\A)$ just in terms of vector bundles
or in other words $D9$- and $\overline {D9}$-brane configurations.
Now we see how to describe this restriction in a way that has broader
validity: the ninebranes correspond to projective modules, and the definition
of $K(X)$ in terms of ninebranes is a special case of the definition of
$K(\A)$ for any ring $\A$ in terms of projective modules.

Now let us drop the assumption that $\A$ is commutative (but keep 
associativity).
If $\A$ is a noncommutative ring, one can canonically associate with it
a second ring $\A^{op}$ (called the opposite ring)
whose elements are in one to one correspondence
with elements of $\A$, but which are multiplied in the opposite order.
Thus, for every element $a\in \A$, there is a corresponding $a^{op}\in
\A^{op}$, with the multiplication law of $\A^{op}$ being
$a^{op}b^{op}=(ba)^{op}$.

For a noncommutative ring, we must distinguish two types of modules,
left-modules and right-modules.  $\A$ acts on the left on a left-module
$M$, the basic axiom being that for $a,a'\in \A$ and $m\in M$, one has
$(aa')(m)=a(a'(m))$.  $\A$ acts on the right on a right-module $M$,
the basic axiom being that $m\cdot (aa')=(m\cdot a)\cdot a'$.  A left-module
of $\A$ is the same as a right-module of $\A^{op}$, and vice-versa.

The definition of projective module is the same as in the commutative
case: a left or right $\A$-module
$M$ is projective if there is another left or right 
$\A$-module $N$ such that $M\oplus N$ is
a free left  or right $\A$-module.  From the semigroup of projective left
$\A$-modules, we form a group called $K(\A)$.  We can also form a semigroup
from the projective right $\A$-modules; it is the same as the semigroup
of projective left $\A^{op}$-modules, so the associated abelian group
is  $K(\A^{op})$.
For the rings we are interested in, there is generally an operation of
complex (or hermitian) conjugation that maps $\A$ to $\A^{op}$.
  Consequently, the $K$-groups formed using
left or right $\A$-modules are equivalent to each other under complex 
conjugation.

\def\B{{\cal B}}
Given two rings $\A$ and $\B$,  a bimodule (or $(\A,\B)$-bimodule)
is a group $M$ which is
simultaneously a left $\A$-module and a right $\B$-module, with the left
action of $\A$ and the     right action of $\B$ commuting.
For $a\in \A$, $b\in \B$, and $m\in M$, the left action of $\A$ is
denoted $a(m)=am$, the right action of $\B$ is denoted $b(m)=mb$, and
since we assume that $\A$ and $\B$ commute, we have $(am)b=a(mb)$;
we abbreviate these expressions as $amb$.

We can now define a very important relationship between rings, called
Morita equivalence.  It was first exploited in connection with $D$-branes
in \cds; for further developments see \ref\schw{A. Schwarz,
``Morita Equivalence And Duality,'' Nucl. PHys. {\bf 534} (1998) 720, 
hep-th/9805034, A. Schwarz and B. Pioline, ``Morita Equivalence And
$T$ Duality (Or $B$ Versus Theta),'' JHEP {\bf 9908:021} (1999), 
hep-th/9908019.}.
  Let $M$ be an $(\A,\B)$-bimodule.  The choice of $M$ enables
us to define a map from left $\B$-modules to left $\A$-modules:
given a left $\B$-module $W$, we map it to the left $\A$-module
$W'=M\otimes_{\B}W$.  Similarly, if we are given a $(\B,\A)$-bimodule $N$,
we can map a left $\A$-module $W'$ to a left $\B$-module $W''$ by
$W''=N\otimes_{\cal A}W'$.  If these two operations are inverse to each
other, in other words if one has $W''=W$ for all $W$, we say that $\A$ and
$\B$ are Morita-equivalent.  In this case, the semigroup of left 
$\A$-modules is isomorphic to the semigroup of left $\B$-modules, 
and $K(\A)=K(\B)$.

The meaning of this relationship in open string field theory was
explained in \ref\sw{N. Seiberg and E. Witten, ``String Theory
And Noncommutative Geometry,'' JHEP {\bf 99090:032} (1999), hep-th/9908142},
section 6.4.  
For any open string boundary condition $\alpha$,
one defines a ring $\A_{\alpha}$ consisting of the $\alpha$-$\alpha$
open strings.  For any other boundary condition $\gamma$, the 
$\alpha$-$\gamma$ open strings form a left $\A_\alpha$-module 
$M_{\alpha\gamma}$,  and the $\gamma$-$\alpha$ open strings form a 
right $\A_\alpha$-module $M_{\gamma\alpha}$.  
Presumably these are in a suitable sense
a basis of left and right $\A_\alpha$-modules.
Now consider any other
open string boundary condition $\beta$ and the associated algebra 
$\A_\beta$.  One forms left and right $\A_\beta$ modules $M_{\beta\gamma}$
and $M_{\gamma\beta}$ in the same way, using $\beta$-$\gamma$ and
$\gamma$-$\beta$ open strings.  Obviously, there is a natural association
between left $\A_\alpha$ modules and left $\A_\beta$ modules, namely
\eqn\mumbo{M_{\alpha\gamma}\leftrightarrow M_{\beta\gamma}.}
So $K(\A_\alpha)=K(\A_\beta)$ for all $\alpha$ and $\beta$.  We established
this last fact directly without talking about Morita equivalence, but in
fact, it seems that the relation between $\A_\alpha$ and $\A_\beta$ is that
they are Morita-equivalent.  Indeed, one has natural $(\A_\alpha,\A_\beta)$
and $(\A_\beta,\A_\alpha)$ bimodules $M_{\alpha\beta}$ and $M_{\beta\alpha}$
constructed from the $\alpha$-$\beta$ and $\beta$-$\alpha$ open strings;
it seems very likely that these bimodules have the right properties to
establish a Morita equivalence between $\A_\alpha$ and $\A_\beta$.

Now let us consider a more elementary version of this, with the goal
of adding some details to the discussion in section 3.2.  Let
$M_n$ be the algebra of $n\times n$ complex matrices.  The most
obvious left $M_n$-module is the space $W$ of $n$-component column vectors.
This is a projective module because, as a left module, $M_n$ consists
of $n$ copies of $W$ ($M_n$ consists of $n\times n$ matrices; in the left
action of $M_n$ on itself, $M_n$ acts separately on each of the $n$ columns
making up the $n\times n$ matrix, so these comprise $n$ copies of $W$).  
So if $W'$ is the sum of $n-1$
copies of $W$, then $W\oplus W'$ is free, and thus $W$ is projective.
It can be shown that every projective 
left $M_n$ module is the sum of copies
of $W$.  So the semigroup of projective left $M_n$ modules is isomorphic
to the semigroup of non-negative integers, and $K(M_n)=\Z$.

Since $K(M_n)$ is thus independent of $n$, one might wonder if the $M_n$'s
of different $n$ are Morita-equivalent.  This is indeed so.
For any positive integers $n$ and $k$, the  most obvious $(M_n,M_k)$ bimodule
is the space $C_{n,k}$ of $n\times k$ matrices, with $M_n$ acting on the left
and $M_k$ on the right.  Likewise, the most obvious $(M_k,M_n)$
bimodule is $C_{k,n}$.  It can be shown that these
 bimodules establish the Morita equivalence
between $M_n$ and $M_k$.

Now, we reconsider the torsion $H$-field that was discussed in section 3.2.
For any twisted bundle $W$, there is an algebra $\A_W(X)$ generated
by the ground states of the $W$-$W$ open strings.  An element of
$\A_W(X)$ is a section of $W\otimes \bar W$.  Given any two different twisted
bundles $W$ and $W'$, we can try to make a Morita equivalence between
$\A_W(X)$ and $\A_{W'}(X)$ 
by specializing to this situation the more abstract discussion
of general open string algebras $\A_\alpha$ and $\A_\beta$ that was given above.
This means that we should find $(\A_W(X),\A_{W'}(X))$ and $(\A_{W'}(X),\A_W(X))$
bimodules by looking at the ground states of the 
$W$-$W'$ and $W'$-$W$ open strings.  The ground states of the $W$-$W'$ open
strings are sections of $W\otimes \overline{W'}$; the space of these sections
is a $(\A_W(X),\A_{W'}(X))$ bimodule $M$.  Likewise, the space of ground
states of the $W'$-$W$ open strings is a $(\A_{W'}(X),\A_W(X))$ bimodule  $N$
consisting of the sections of $W'\otimes \bar W$.  The bimodules
$M$ and $N$ do indeed establish a Morita equivalence between $\A_W(X)$
and $\A_{W'}(X)$.  Indeed, in proving this one can work locally on $X$.
Locally, $W$ and $W'$ are trivial, and (if $n$ and $k$ are the ranks
of $W$ and $W'$) we have locally as noted in section 3.2,
 $\A_W(X)=\A(X)\otimes M_n$,
$\A_{W'}(X)=\A(X)\otimes M_k$.  The same computation that gives a Morita
equivalence of $M_n$ with $M_k$ shows that $\A_W(X)$ is Morita
equivalent to $\A_{W'}(X)$.

A projective right $\A_W(X)$ module is the space of sections of
$E\otimes \bar W$, for any twisted bundle $E$.  (If $s^\alpha{}_j$
is such a section -- where $\alpha$ is an $E$ index and $j$ a $\overline
W$ index -- and $w^k{}_l$ is a section of $W\otimes \bar W$, then the product of $s$
and $w$ is defined by $(sw)^\alpha{}_j=\sum_ks^\alpha{}_kw^k{}_j$.)
So projective right $\A_W(X)$ modules are in natural
one-to-one correspondence with
twisted bundles $E$, and the $K$-group of projective right $\A_W(X)$ modules
is the group of pairs $(E,F)$, subject to the usual equivalence relation. 
The Morita equivalence of $\A_W$ and $\A_{W'}$, for any two twisted bundles
$W$ and $W'$ with the same $H$,
is just a fancy way of describing the natural map
between  $\A_W$-modules and $\A_{W'}$-modules that comes from the correspondence
\eqn\olpo{E\otimes \overline W \leftrightarrow E\otimes \overline{W'}.}
This one-to-one map between right $\A_W$-modules and right $\A_{W'}$-modules
(or the corresponding map for left modules) leads  to the conclusion
that $K(\A_W)=K(\A_{W'})$.

\bigskip
This work was supported in part by NSF Grant PHY-9513835 and the Caltech
Discovery Fund.  I am grateful to K. Hori and G. B. Segal for  comments
and explanations.
\listrefs
\end